\documentstyle[aps,12pt]{revtex}

\begin{document}
\draft
\author{O. B. Zaslavskii}
\address{Department of Physics, Kharkov State University, Svobody Sq.4, Kharkov\\
310077, Ukraine\\
E-mail: aptm@kharkov.ua}
\title{Two-dimensional dilaton gravity and spacetimes with finite curvature at the
horizon away from the Hawking temperature}
\maketitle

\begin{abstract}
It is shown that static solutions with a finite curvature at the horizon may
exist in dilaton gravity at temperatures $T\neq T_{H}$ (including $T=0$)
where $T_{H}\,$is the Hawking one. Hawking radiation is absent and the state
of a system represents thermal excitation over the Boulware vacuum. The
horizon remains unattainable for a observer because of thermal divergences
in the stress-energy of quantum fields there. However, the curvature at the
horizon is finite, when measured from outside, since these divergences are
compensated by those in gradients of a dilaton field. Spacetimes under
consideration are geodesically incomplete and the coupling between dilaton
and gravity diverges at the horizon, so we have ''singularity without
singularity''.
\end{abstract}


In recent years a great interest focused on two-dimensional (2D) theories of
dilaton gravity. One of the reason of this consists in the possibility to
trace in detail the process of black hole evaporation since on the
semiclassical level spacetime evolution is described by differential
equations directly following from the Lagrangian even if quantum effects are
taken into account \cite{callan}. It turned out that such a kind of theories
possesses a rather rich set of exactly solvable models. These circumstances
give us hope to get insight into subtle features of black hole evolution.
Especially important is the question whether final geometry can be regular
or not. In Ref. \cite{bose} (BPP) the exactly solvable model was proposed in
which, under certain choice of parameters, a black hole leaves after
evaporation everywhere regular geodesically complete geometry which
infinitely extends to a region of strong coupling (''semi-infinite''
throat). In so doing, quantum fields are in the radiationless state in which
effects of vacuum polarization tend to zero at Minkowski infinity. In Ref. 
\cite{cruz} (CN) more general model was suggested interpolating between the
BPP and RST \cite{rst} ones. The qualitatively new feature of the CN model
consists in that, although a static radiationless geometry with curvature
finite everywhere is still possible, it becomes geodesically incomplete.

Meanwhile, the CN model possesses some other intriguing features which were
not noticed in \cite{cruz}. Namely, we will show below that at the boundary
of spacetime $g_{00}\rightarrow 0$, so it represents a Killing horizon. At
this point stresses of quantum fields due to back reaction diverge since the
solution under discussion is radiationless (temperature $T=0$) whereas the
usual condition of their finiteness demands $T=T_{H}$ where $T_{H}$ is the
Hawking temperature (see, for instance,\cite{paul}). Thus, the intimate
connection between regularity of a geometry on a horizon and equality of
temperatures $T=T_{H}$ ceases to exist and we have a static geometry with a
finite curvature at the horizon at $T\neq T_{H}$ and the infinite quantum
stresses! In this respect the significance of the CN model is beyond the
context of concrete problems of 2D dilaton gravity: relative simplicity of
the 2D case enables us in the best possible way to review unbiasedly some
fundamentals of black hole physics and elucidate opportunities which
probably may occur in a 4D world but remain hidden because of complexity of
the situation in the latter case.

In the present paper we argue also that there exists a whole range of
temperatures for which properties of spacetime sketched above hold true. We
exploit exactly solvable models of 2D gravity \cite{kaz}, \cite{zasl99}
(with respect to which the CN one is the particular case) and show that the
results remain valid for a whole class of such theories provided the
coupling between gravitation and dilaton obeys some general restrictions.

Let us consider the system described by the action 
\begin{equation}
I=I_{0}+I_{PL}  \label{1}
\end{equation}
where 
\begin{equation}
I_{0}=\frac{1}{2\pi }\int_{M}d^{2}x\sqrt{-g}[F(\phi )R+V(\phi )(\nabla \phi
)^{2}+U(\phi )]\text{,}  \label{2}
\end{equation}
$I_{PL\text{ }}$is the Polyakov-Liouville action \cite{polyakov}
incorporating effects of Hawking radiation and its back reaction on the
background metric for a multiplet of N scalar fields, and the boundary terms
omitted. In what follows we will use the quantities $T_{\mu \nu }$ defined
according to 
\begin{equation}
\delta I=\frac{1}{2}\int d^{2}x\sqrt{\left| g\right| }\delta g^{\mu \nu
}T_{\mu \nu }\text{,}  \label{t}
\end{equation}
so the filed equations which are obtained by varying a metric have the form $%
T_{\mu \nu }=0$.

In the conformal gauge 
\begin{equation}
ds^{2}=-e^{2\rho }dx_{+}dx_{-}  \label{m}
\end{equation}
the Polyakov-Liouville action reads $I_{PL}=-\frac{2\kappa }{\pi }\int
d^{2}x\partial _{+}\rho \partial _{-}\rho =\frac{\kappa }{\pi }\int d^{2}x%
\sqrt{g}(\nabla \rho )^{2}$. After integration by parts the action (\ref{1})
is reduced (up to boundary terms) to $I=\frac{1}{2\pi }\int d^{2}x\sqrt{g}%
[V(\nabla \phi )^{2}+2\nabla \rho \nabla F+2\kappa (\nabla \rho
)^{2}+4\lambda ^{2}e^{\eta }]$, where by definition $\eta =\int d\phi \omega 
$, $\omega =U^{\prime }/U$. In what follows we restrict ourselves by exactly
solvable models of dilaton gravity \cite{kaz}, \cite{zasl99} that implies
the constraint on the form of the potential $V=\omega (u-\frac{\kappa \omega 
}{2})+\gamma (u-\kappa \omega )^{2}$ where $u=F^{\prime }$. In what follows
we consider only the simplest choice $\gamma =0$, so 
\begin{equation}
V=\omega (u-\frac{\kappa \omega }{2})  \label{v}
\end{equation}
Introducing new fields $\Omega $ and $\chi $ instead of $\phi $ and $\rho $
according to $\tilde{F}\equiv F-\kappa \eta =2\kappa \Omega $, $\eta =2(\chi
-\Omega -\rho )$, after simple rearrangement we obtain $I=\frac{1}{\pi }\int
d^{2}x\sqrt{g}\{\kappa [(\nabla \chi )^{2}-(\nabla \Omega )^{2}+2\lambda
^{2}e^{2(\chi -\Omega -\rho )}\}.$ Corresponding equations of motion have
the form $2\kappa \partial _{+}\partial _{-}\Omega =-\lambda ^{2}e^{2(\chi
-\Omega )}$, $2\kappa \partial _{+}\partial _{-}\chi =-\lambda ^{2}e^{2(\chi
-\Omega )}$. One can choose the gauge $\chi =\Omega $ whence $\partial
_{+}\partial _{-}\tilde{F}=-\lambda ^{2}$. This equation should be
supplemented by the constraint equations $T_{++}=T_{--}=0$ ($T_{\mu \nu }=2%
\frac{\delta I}{\delta g^{\mu \nu }}$). The expressions for the classical
parts of $T_{++\text{ }}$ and $T_{--}$ follow directly from the general
expression for covariant components 
\begin{equation}
T_{\mu \nu }^{(0)}=\frac{1}{2\pi }\{2(g_{\mu \nu }\Box F-\nabla _{\mu
}\nabla _{\nu }F)-Ug_{\mu \nu }+2V\nabla _{\mu }\phi \nabla _{\nu }\phi
-g_{\mu \nu }V(\nabla \phi )^{2}\}\text{,}  \label{fcl}
\end{equation}
which can be obtained by varying the action $I_{0}$. The formula for the
quantum contribution can be obtained from the conservation law and conformal
anomaly and has the form 
\begin{equation}
T_{\pm \pm }^{(PL)}=\frac{2\kappa }{\pi }[(\partial _{\pm }\rho
)^{2}-\partial _{\pm }^{2}\rho +t_{\pm }]  \label{stress}
\end{equation}
where the function $t_{\pm }(x_{\pm })$ are determined by the boundary
conditions. From (\ref{fcl}) and (\ref{stress}) we have the equation $%
\partial _{\pm }^{2}\tilde{F}=2\kappa t_{\pm }$. To find the explicit form
of $t_{\pm }$, let us impose the condition that at the right infinity
quantum fields should be at a finite temperature $T$. Then in asymptotically
flat coordinates $\sigma _{\pm }=t\pm \sigma $ connected with $x_{\pm }$
according to $\lambda x_{\pm }=\pm e^{\pm \lambda x_{\pm }}$ stresses take
the form $T_{\pm \pm }^{(\sigma )}=-\pi T^{2}/12$ at $\sigma \rightarrow
\infty $ where $ds^{2}=-d\sigma _{+}d\sigma _{-}$. On the other hand,
asymptotically $2\rho \approx -\ln (-\lambda ^{2}x_{+}x_{-})$ to match the
region of a linear dilaton vacuum $\phi =-\sigma \lambda $. Substituting
this into (\ref{stress}) and performing transformation of tensor components
between two coordinate systems, we have $t_{\pm }=\frac{1}{4}x_{\pm }^{-2}(1-%
\frac{T^{2}}{T_{0}^{2}})$. Then, integrating the equations of motion for $%
\tilde{F}$ we obtain 
\begin{eqnarray}
ds^{2} &=&g(-dt^{2}+d\sigma ^{2})\text{, }g=\exp (2\rho +2y)\text{, }%
y=\lambda \sigma \text{, }2\rho =-\int \omega d\phi \text{.}  \label{bas} \\
\tilde{F}(\phi ) &=&f(y)\equiv e^{2y}-By+C\text{, }B=\kappa
(1-T^{2}/T_{0}^{2})\text{, }T_{0}=\lambda /2\pi \text{.}  \nonumber
\end{eqnarray}

The expressions (\ref{bas}) describe generic static configurations
gravitational and dilaton fields in exactly solvable models of dilaton
gravity at finite temperatures. Hereafter we concentrate ourselves on $%
\tilde{F}(\phi )$ such that $\tilde{F}(\phi )$ has one simple minimum at
some real $\phi _{0}$ where $\tilde{F}^{\prime }(\phi _{0})=0$. The well
known example is the RST model \cite{rst}. Let the coefficient $B>0$. The
structure of spacetime depends crucially on the relationship between $%
f_{\min }=f(y_{0})$ (the minimum value of $f(y)$ achieved in the point $%
y_{0} $) and $\tilde{F}(\phi _{0})$. It follows from (\ref{bas}) that $\frac{%
d\phi }{dy}=\frac{f^{\prime }(y)}{\tilde{F}^{^{\prime }}(\phi )}$. If $%
f_{\min }<$ $\tilde{F}(\phi _{0})$, the dilaton value changes monotonically
from $\phi =-\infty $ at right infinity $y=\infty $ to $\phi =\phi _{0}$
where the spacetime is singular and cannot be continued further. For the RST
model the structure of spacetime at the equilibrium temperature $T=T_{0}$ is
analyzed in detail in \cite{solod} (see also generalization in \cite{zasl99}%
). If $f_{\min }>\tilde{F}(\phi _{0})$, the dilaton field takes its values
in the limits $(-\infty $, $\phi _{1}$) where $\phi _{1}<\phi _{0}$, $\tilde{%
F}^{\prime }(\phi )$ changes its sign nowhere.

In what follows we dwell upon on the special case $\tilde{F}(\phi
_{0})=f_{\min }$, so $C=C^{*}$ where 
\begin{equation}
C^{*}=\tilde{F}(\phi _{0})+C_{0}\text{, }C_{0}=-\frac{B}{2}(1+\ln \frac{2}{B}%
)  \label{c}
\end{equation}
Then $f^{\prime }(y)$ and $\tilde{F}^{\prime }(\phi )$ turn into zero
simultaneously and $\frac{d\phi }{dy}$ does not changes its sign, so the
dependence $\phi (y)$ is monotonic. As a result, $\phi \rightarrow \infty $
when $y\rightarrow -\infty $, i.e. at the event horizon. In the vicinity of
the point $y_{0}$ one can use the power expansion, so the equality $\tilde{F}%
(\phi )=f(y)$ turns into $\frac{\tilde{F}^{\prime \prime }(\phi _{0})}{2}%
(\phi -\phi _{0})^{2}+\frac{\tilde{F}^{\prime \prime \prime }(\phi _{0})^{3}%
}{6}(\phi -\phi _{0})^{3}+...=\frac{f^{\prime \prime }(y_{0})}{2}%
(y-y_{0})^{2}+\frac{f^{\prime \prime \prime }(y_{0})}{6}(y-y_{0})^{3}+...$
whence it is clear that the Riemann curvature $R=-\lambda ^{2}g^{-1}\frac{d}{%
dy}(g^{-1}\frac{dg}{dy})=-2\lambda ^{2}\frac{d^{2}\phi }{dy^{2}}e^{-2\phi
-2y}$ is finite in the point $y_{0}$.

Let us consider the concrete example. We assume that the functions $f(y)$
and $\tilde{F}(\phi )$ obey the conditions described above which guarantee
the absence of the singularity outside the horizon. We do not specify the
exact form of $\tilde{F}(\phi )$ in the whole region and only assume that in
the vicinity of the horizon it reads 
\begin{equation}
\tilde{F}\approx e^{-2\phi }+b\phi \text{, }\phi \rightarrow \infty \text{.}
\label{ex}
\end{equation}
The condition (\ref{v}) of exact solvability implies that $V\approx
(b-2e^{-2\phi })\omega +\frac{\kappa \omega ^{2}}{2}$ near the horizon. We
find that at $y\rightarrow -\infty $ 
\begin{equation}
\phi \rightarrow \infty \text{, }g\sim e^{2\left| y\right| (B/b-1)},R\sim
e^{-2\left| y\right| (2B/b-1)}  \label{asym}
\end{equation}
Thus, if $B/b<1$ the surface $y=-\infty $ represent a horizon where $g=0$.
In the Schwarzschild gauge $x=\int d\sigma g$, $ds^{2}=-gdt^{2}+g^{-1}dx^{2}$
we have an usual behavior $g\sim x-x_{h}$ near the horizon located at $%
x=x_{h}$. If $2B/b\geq 1$, the Riemann curvature on this surface is finite.
It is remarkable that both conditions are consistent with each other. Thus,
if $\frac{1}{2}\leq \frac{B}{b}<1$ or, equivalently, 
\begin{equation}
1-\frac{b}{\kappa }<\frac{T^{2}}{T_{0}^{2}}\leq 1-\frac{b}{2\kappa }
\label{cond}
\end{equation}
we have a black hole with the Riemann curvature finite everywhere including
the horizon. This inequality is not satisfied by $T=T_{0}$ when, as is well
known, a horizon is regular (the usual Hartle-Hawking state). However, there
is no contradiction here since at $T=T_{0}$ the coefficient $B=0$ and
asymptotic behavior of such a solution \cite{zasl99}, \cite{solod} has
nothing to do with (\ref{asym}). Thus, according to (\ref{cond}), the
Hartle-Hawking state is not included in the set of states under discussion
and there are two possibilities to have the horizon with a finite curvature
which cannot match continuously: $T=T_{0}$ (the isolated point) or $T$ obeys
the inequality (\ref{cond}) (the whole range of temperatures).

In the particular case of the RST\ model $b=\kappa $, so (\ref{cond}) is
reduced, according to (\ref{bas}), to the condition 
\begin{equation}
0<T\leq \frac{T_{0}}{\sqrt{2}}  \label{temp}
\end{equation}
The case $T=0$ is now trivial since for solutions under discussion $C=0$
according to (\ref{c}), eq. (\ref{bas}) gives us $e^{-2\phi }+\kappa \phi
=e^{2y}-\kappa y$ and, with the condition of asymptotical flatness $\phi
(\infty )=-\infty $, we have a linear dilaton vacuum solution $\phi =-y$
with a flat spacetime. Therefore, for the RST model a horizon with a finite
curvature may exist only at nonzero temperature. However, for a generic $%
\kappa <b\leq 2\kappa $ it is possible for such a horizon to exist even at $%
T=0$.

The case $T=0$ is tractable for more detailed investigation. Let us consider
the example 
\begin{equation}
\tilde{F}=e^{2\gamma (\phi )}-\kappa \gamma (\phi )  \label{t=0}
\end{equation}
Then, taking into account that for this model the quantity $C^{*}$ depends
on temperature according to $C^{*}=C_{0}(0)-C_{0}(T)$, so $C^{*}=0$ at $T=0$%
, we find from (\ref{bas}) the solution $y=\gamma (\phi )$. Let us choose,
for instance, 
\begin{equation}
\gamma =-\phi -\frac{1}{2}\ln (1+e^{2\phi })  \label{g}
\end{equation}
Then $g=(1+e^{2\phi })^{-1}$. After simple calculations we find the Riemann
curvature $R=4(1-g)(g-2)^{-3}$. At the horizon $\phi =\infty $, $g=0$ the
curvature remains finite in spite of quantum stresses diverge there. For
instance, $T_{y}^{y(PL)}=-\frac{\lambda ^{2}}{24}(1+\frac{1}{2}e^{-2\phi
})^{-2}g^{-1}$.

Thus, we arrive at a rather surprising conclusion: spacetimes with a finite
curvature at the horizon may exist at a temperature different from the
Hawking one and, moreover, their temperature may be equal to zero! This
conclusion is the main result of the present paper.

What is the physical reason for such an unusual behavior? It is instructive
to look at fields equations $T_{\mu }^{(0)\nu }+T_{\mu }^{(PL)\nu }=0$. The
classical part of the effective stress-energy tensor $T_{\mu }^{(0)\nu }$
contains terms with gradients of the dilaton field $\phi $. For instance, $%
T_{x}^{(0)x}=(2\pi )^{-1}[V(\nabla \phi )^{2}-U+2\nabla ^{0}\nabla _{0}F]$.
A simple estimate based on (\ref{ex}), (\ref{asym}) shows that near the
horizon where $y\rightarrow -\infty $ and $\phi \approx -\frac{B}{b}y$, $%
T_{x}^{(0x1}\sim g^{-1}\rightarrow \infty $. On the other hand, in the
Scwharzschild coordinates we have $T_{x}^{(PL)x}=-\frac{\pi }{6}[T^{2}-\frac{%
1}{4\pi }(\frac{dg}{dx})^{2}]g^{-1}\sim g^{-1}\rightarrow \infty $ if $T\neq
T_{H}$ \cite{paul}. Thus, each part of $T_{\mu }^{\nu }$ diverges separately
and a regular metric near the horizon arises as a result of mutual
compensation of these divergences. Meanwhile, the usual proof of the fact
that $T=T_{H}$ relies strongly upon the regularity of the stress-energy
tensor of quantum field at the horizon. The corresponding criteria \cite
{paul} appeal to the behavior of $T_{\mu }^{(PL)\nu }$ but do not take into
account properties of $T_{\mu }^{(0)\nu }$, i.e. they were considered in
isolation from the dynamic contents of the theory whose solution a given
metric represents. Usually, such an approach is justified since a classical
part of field equation is regular, as was tacitly assumed in \cite{paul},
but the present case is exceptional in that either quantum or classical
contributions to field equations diverge near the horizon.

Thus, the curvature may be finite everywhere even when the stress-energy
tensor describing the back reaction of quantum fields diverges on the
horizon. The situation is sharply contrasted with that in general relativity
where the stress-energy tensor singular at a horizon is inconsistent with
the regularity of a metric and of the Einstein tensor. The reason consists
in that now we have, apart from a metric, one more classical field - the
dilaton one coupled to a metric. Gradients of this field compensate the
divergencies due to back reaction of quantum fields.

It can be readily seen from the above formulas that the proper distance
between the horizon and any other point $l=\lambda \int_{-\infty }^{y}dy%
\sqrt{g}$ is finite. Therefore, the spacetime region outside the horizon is
geodesically incomplete. The dilaton field $\phi $ cannot be continued
further across the horizon since $\phi $ diverges there. The situation with
continuation across such a surface leads, in general, to complex dilaton
values \cite{accel}. To avoid complex dilaton field, one may try to redefine
the dilaton field choosing, say, $x(\phi )$ instead of $\phi $ where $x$ is
a Schwarzschild-like coordinate. However, any redefinition cannot abolish
the fact that the quantity $\tilde{F}$ describing the coupling between the
dilaton and curvature diverges at the horizon. There is more deep reason to
believe that any attempts of continuation across the horizon for our
solutions should be rejected. The point is that any observer approaching the
horizon from outside moves along a non-geodesic path and perceives fluxes of
quantum field which become infinite at the horizon. Therefore, the horizon
remains in fact unaccessible. The surface $\phi =\infty $ acts as a boundary
of spacetime and the usual criteria of geodesic completeness is physically
irrelevant for the solution under discussion. Of course, nothing prevents
making a formal analytic continuation of the metric itself but, if the
dynamic contents of the theory (interaction between gravitational, dilaton
and quantum scalar fields) is taken into account, the region inside a
horizon seems not to have a physical meaning for the model in question - at
least, if one starts from the outside region. Thus, actually the loss of
information inside the horizon is much more severe than in an usual case
where at least an observer brave enough to dive into a black hole can get
information about the region inside the horizon. In fact the notion of an
event horizon changes its ordinary meaning since there are no events inside
a horizon at all.

It is seen from (\ref{ex}) that $\tilde{F}$ diverges at the horizon where $%
\phi =\infty $ and so does the total entropy including the contribution of
quantum fields which for exactly solvable models is proportional (up to a
constant) to the horizon value of this function\cite{zasl99}. These
divergences have the same nature as those in the Schwarzschild background
where entropy of quantum fields is infinite if $T\neq T_{H}$. The difference
between these two situations manifests itself, however, in that for a
black-hole metric in general relativity disparity between two temperatures
makes a horizon singular and in fact destroys it completely whereas in our
case the horizon remains regular in the sense that the Riemann curvature
remains finite there.

The features of spacetime discussed above are shared by a whole classes of
exactly solvable models of dilaton gravity for which (i) $\tilde{F}^{\prime
}(\phi )$ has a simple zero at some $\phi _{0}$, (ii) $\tilde{F}\sim \phi $
at $\phi \rightarrow \infty $, (iii) the constant in the solution is chosen $%
C=C^{*}$ in accordance with (\ref{c}). These features, however, have nothing
to do, for example, with the BPP model for which $\tilde{F}=e^{-2\phi }$ and
the choice $C=C^{*}$ leads to a regular semi-infinite ''throat'' \cite{bose}
(generalization to a finite temperature is considered in \cite{static}). The
fact that unusual properties of the solution under discussion are intimately
connected with quantum terms in the action reveals itself in the eq. (\ref
{bas}) and its classical limit directly. Indeed, it follows from (\ref{bas}%
), (\ref{ex}) that in the limit $\kappa =0$ the point $\phi _{0}$ where $%
\tilde{F}^{\prime }=0$ moves to $\phi _{0}=\infty $, the coefficients $%
B=0=C^{*}$, $b\sim \kappa =0$ and eq. (\ref{bas}) with $C=C^{*}$ turns into $%
\tilde{F}=e^{-2\phi }=e^{2y}$ whose solution is a linear dilaton vacuum $%
\phi =-y$.

Usually, the temperature of Hawking radiation in a black hole background is
determined by characteristics of a spacetime (say, an event horizon radius
of a Schwarzschild hole). The fact that temperature may be arbitrary
actually means that Hawking radiation as such is absent, so we deal with
thermal excitation over the Boulware state which can be treated at any
temperature. In particular, the choice $T=0$ corresponds to the Boulware
vacuum. Therefore, the type of solution considered in the present paper
might shed light to the fate of a black hole after evaporation, being a
candidate on the role of a ''remnant''. This, however, needs further
treatment based on a dynamic scenarios and is beyond the scope of the
present paper.

While usually the notion of singularity implies that it is geometry which
exhibits singular behavior, in our case divergencies manifest themselves in
the dynamic characteristics (components of the stress-energy tensor) and in
the coupling between the dilaton filed and curvature. These divergences
along with the geodesic incompleteness of the metric in fact mean that the
solution in question is singular in spite of the metric itself is regular -
at least, in the restricted sense: as one approaches the horizon from
outside, the Riemann curvature remains finite. So, we have ''singularity
without singularity''.

In fact, we are faced with a qualitatively new type of objects in
gravitation which occupies an intermediate place between regular black holes
and naked singularities. It is of interest to elucidate whether coupling
between gravitation and dilaton or other classical fields can produce the
same type of solutions in the four-dimensional case.

I am grateful to Sergey Solodukhin for fruitful discussion and to Alessandro
Fabbri for helpful correspondence.





%
%

\end{document}